\documentclass[aip,apl, preprint, numerical]{revtex4-1}

\usepackage[version=3]{mhchem} 
\usepackage[T1]{fontenc}       
\usepackage{graphicx}          
\usepackage{epstopdf}
\usepackage{dcolumn}           
\usepackage{bm}                
\usepackage{color}
\usepackage{ulem}

\begin{document}
\title{Two-dimensional optical plasmons with mixed polarization on anisotropic resonant metasurface}

\author{Anton Samusev}
\email{a.samusev@metalab.ifmo.ru}
\affiliation{Department of Nanophotonics and Metamaterials, ITMO University, St. Petersburg 197101, Russia}
\author{Ivan Mukhin}
\affiliation{Department of Nanophotonics and Metamaterials, ITMO University, St. Petersburg 197101, Russia} 
\affiliation{St. Petersburg Academic University, St. Petersburg 194021, Russia}
\author{Radu Malureanu}
\affiliation{Department of Photonics Engineering, Technical University of Denmark, 2800 Kongens Lyngby, Denmark}
\affiliation{National Centre for Micro- and Nano-Fabrication, Technical University of Denmark, 2800 Kongens Lyngby, Denmark}
\author{Osamu Takayama}
\affiliation{Department of Photonics Engineering, Technical University of Denmark, 2800 Kongens Lyngby, Denmark}
\author{Dmitry V. Permyakov}
\affiliation{Department of Nanophotonics and Metamaterials, ITMO University, St. Petersburg 197101, Russia}
\author{Ivan S. Sinev}
\affiliation{Department of Nanophotonics and Metamaterials, ITMO University, St. Petersburg 197101, Russia}
\author{Dmitry Baranov}
\affiliation{Department of Nanophotonics and Metamaterials, ITMO University, St. Petersburg 197101, Russia}
\author{Oleh Yermakov}
\affiliation{Department of Nanophotonics and Metamaterials, ITMO University, St. Petersburg 197101, Russia}
\author{Ivan V. Iorsh}
\affiliation{Department of Nanophotonics and Metamaterials, ITMO University, St. Petersburg 197101, Russia}
\author{Andrey A. Bogdanov}
\affiliation{Department of Nanophotonics and Metamaterials, ITMO University, St. Petersburg 197101, Russia}
\author{Andrei V. Lavrinenko}
\affiliation{Department of Nanophotonics and Metamaterials, ITMO University, St. Petersburg 197101, Russia}
\affiliation{Department of Photonics Engineering, Technical University of Denmark, 2800 Kongens Lyngby, Denmark}


\begin{abstract}
Optical metasurfaces have great potential to form the platform for manipulation of surface waves. A plethora of advanced surface-wave phenomena utilizing negative refraction, self-collimation and channeling of 2D waves can be realized through on-demand engineering of dispersion properties of a periodic metasurface. In this letter, we report on the first-time direct experimental polarization-resolved measurement of dispersion of 2D optical plasmons supported by an anisotropic metasurface. We demonstrate that a subdiffractive array of strongly coupled resonant anisotropic plasmonic nanoparticles supports unusual optical surface waves with mixed TE- and TM-like polarizations. With the assistance of numerical simulations we identify dipole and quadrupole dispersion bands. The shape of isofrequency contours changes drastically with frequency exhibiting nontrivial  transformations of their curvature and topology that is consistently confirmed by the experimental data. By revealing polarization degree of freedom for surface waves, our results open new routes for designing of planar on-chip devices for surface photonics.
\end{abstract}

\maketitle

Metasurfaces are a two-dimensional analogue of bulk metamaterials. They represent a dense array (usually  periodic) of subwavelength scatterers,\cite{Pors2014} which are often called meta-atoms. The term {\it metasurface} was introduced recently~\cite{Sievenpiper2003}, but such objects are fairly well-known in electromagnetism as {\it impedance} or \textit{frequency selective surfaces}. They have been actively studied for more than 100 years \cite{Lamb1897} aiming radio frequencies and microwaves.  

Nowadays, metasurfaces are intensively studied in optics and photonics, because they possess many properties of bulk metamaterials being, at the same time, much less lossy, cheaper to fabricate, fully compatible with planar technologies, and ready to be implanted in modern on-chip devices. They offer unprecedented control over phase, amplitude, polarization, propagation directions, and wavefront features of reflected and transmitted waves~\cite{Glybovski2016}. In particular, metasurfaces could essentially increase harvesting of solar energy~\cite{Hedayati2011,Liu2014solar,Yao2014,Akselrod2015} and enhance nonlinear response in optics~\cite{Ding2014,Lee2014,Minovich2015}. The actively developed physics of metasurfaces results in formation of such branches of optics as {\it flat optics} and {\it planar photonics}~\cite{Yu2014, Kildishev2013}. The concept of phase discontinuities, the generalized Snell's and Brewster's laws make metasurfaces very promising for a plethora of applications such as subwavelength focusing and imaging~\cite{Khorasaninejad2016review1,Chen2016, Roy2013, Qin2016, Yuan2014, Taubner2006, Fang2005}, flat lenses~\cite{Yu2014,Verslegers2009, Grbic2008} and holograms~\cite{Zheng2015,Chen2014,Wen2015,Huang2013,Ni2013,Huang2015}, aberration-free, multispectral chiral metalenses, helicity-dependent, and polarization-insensitive lenses~\cite{Aieta2012, Khorasaninejad2016nl, Arbabi2016, Chen2012, Khorasaninejad2016}, light modulators providing efficient control over orbital and spin angular momentum of light~\cite{Zhang2016,Lee2016,Li2013,Liu2016,Bliokh2015,Cardano2015,Devlin2017,Shu2016}.

The majority of the results reported so far are related to free space optics  with functioning of metasurfaces in the transverse configuration. In this case the leaky and quasi-guided resonances resonances play a major role\cite{wood1902xlii,kitson1996full}. However, for on-chip photonic applications, metasurfaces are expected to operate in the in-plane mode, when the surface modes move to the forefront. For routing of optical signals and all-optical networking, the precise control over directivity of surface waves is needed. One of the ways is to involve Dyakonov surfaces waves, since high directivity is their intrinsic feature~\cite{Dyakonov1988,Takayama2009,Takayama2014,Noginov2014}. However, their weak localization and very specific existence conditions impede large-scale implementation of Dyakonov waves in photonic circuits.  

Alternatively, it is possible to exploit the conventional SPPs whose directivity and wave fronts can be engineered via nanostructuring of plasmonic interfaces. So, it was shown in Ref.~\citenum{Liu2013} that isofrequency contours of SPPs  propagating along a metallic grating can have elliptic, flat, or hyperbolic shape depending on geometry of the grating. This results in broadband negative refraction and non-divergent propagation of SPPs. A visible-frequency non-resonant hyperbolic metasurface based on a nanometer-scale silver/air grating was realized in Ref.~\citenum{High2015}.

Recently, it has been predicted  that spectrum of a resonant metasurface described in terms of anisotropic surface conductivity tensor consists of two hybrid TE-TM polarized modes that can be classified as TE-like and TM-like plasmons~\cite{Yermakov2015,Gomez-Diaz2015}. Their isofrequency contours are of elliptic, hyperbolic, 8-shaped, rhombic, or arc form depending on the frequency. Such variety of shapes can support diverse phenomena, e.g. negative refraction, self-collimation, channeling of surface waves, and a giant enhancement of spontaneous emission of quantum emitters due to the large density of optical states. The similar phenomena can be observed for metasurfaces implemented using nanostructured graphene monolayers and naturally anisotropic ultrathin black phosphorus films~\cite{Gomez-Diaz2015,GomezDiaz2016,Correas2016black}. 

In this work, we report the characterization of a resonant anisotropic plasmonic metasurface consisting of a dense array of thin gold elliptic nanoparticles supporting propagation of 2D hybrid plasmons in the near-IR. We characterize  dispersion of both TE- and TM-like plasmons with attenuated total internal reflection spectroscopy and reveal topological transition of their isofrequency contours. 

\section*{Results \label{sec:results}}

\begin{figure}[ht]
\includegraphics[width=1.0\linewidth]{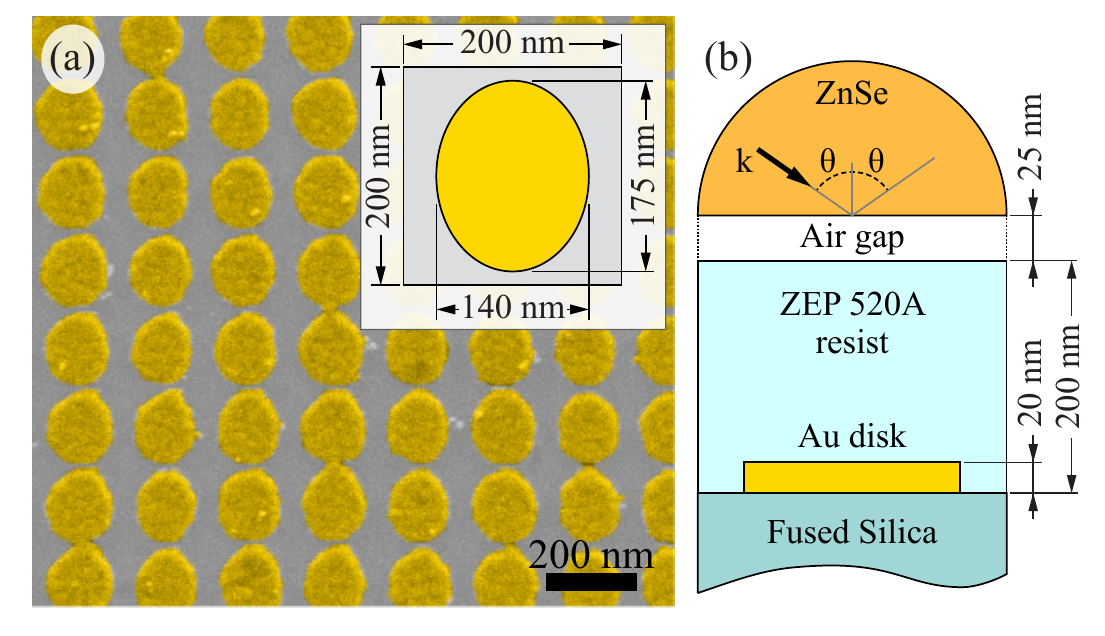}
\caption{
\label{fig:sample_setup}
(a) False color scanning electron micrograph of a small region of the metasurface sample (image taken before sputtering of the cover layer). The inset shows the unit cell used in numerical simulations. (b) Schematic of the experimental geometry for surface waves spectroscopy.
}
\end{figure}

The fabricated anisotropic metasurface was formed by an array of cylindrical gold nanodisks with the elliptical base (see Methods for details of the technological process). The period of the array is 200~nm, while the long and short axes of the nanodisks are 175 and 140~nm, respectively (see Fig.~\ref{fig:sample_setup}a). To facilitate surface waves propagation in the symmetric environment, the sample was subsequently covered by a 200~nm layer of transparent resist (ZEP 520A) with the refractive index closely matching that of silicon oxide in the visible and near-IR spectral regions\cite{Zep520}.

\begin{figure*}[ht]
\includegraphics[width=1.0\linewidth]{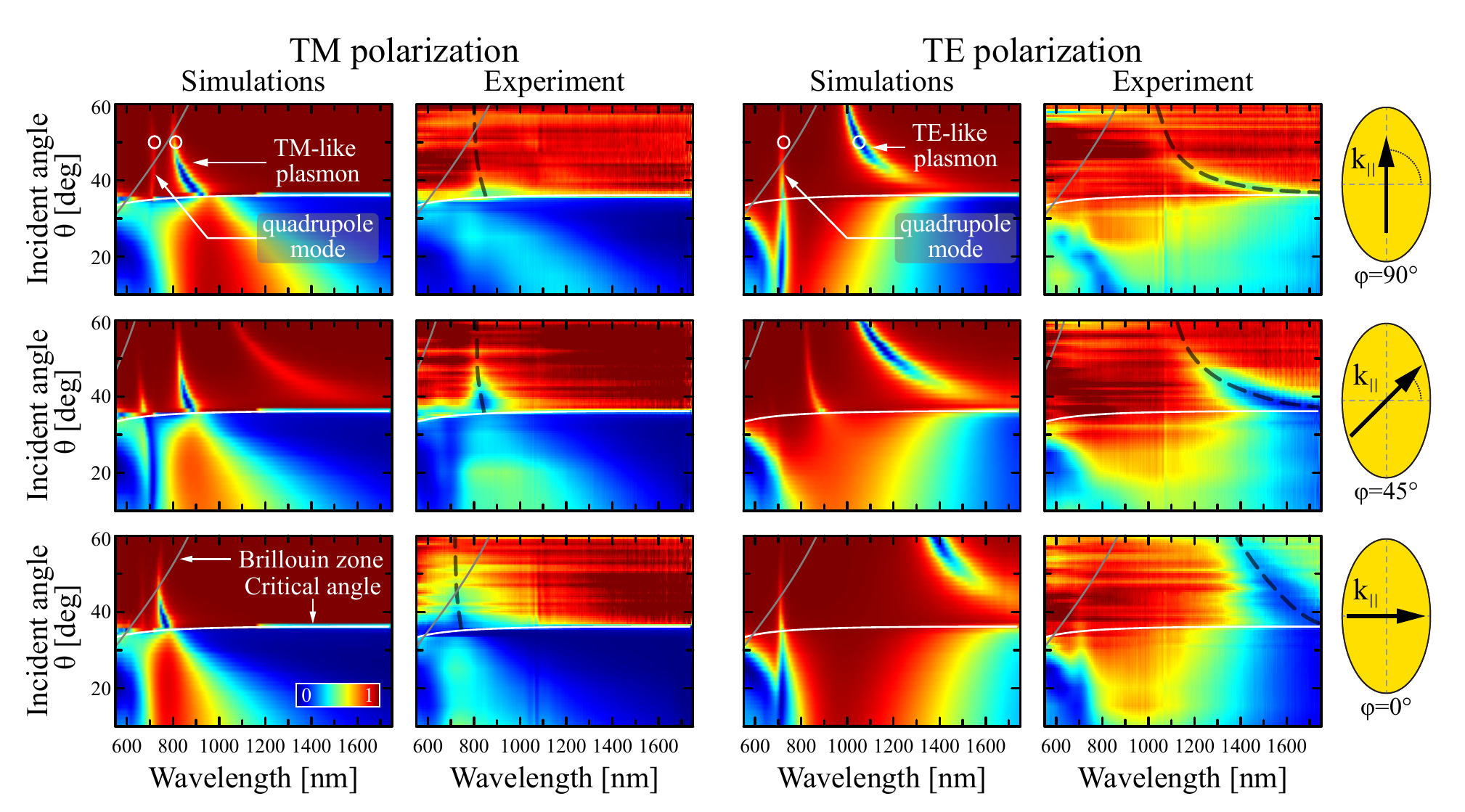}
\caption{
\label{fig:exp_vs_num} 
Measured and simulated angular dependence of reflectance spectra of the anisotropic metasurface coupled to a high-index prism. The data are presented for both TM- and TE-polarized excitation (left and right column couples, respectively).
Top, middle and bottom rows correspond to the plane of incidence forming an angle $\varphi$ of 90$^\circ$, 45$^\circ$, and 0$^\circ$ with the short axis of elliptic particles, as sketched at the right.
The wavelength-dependent critical angle for the ZnSe-resist interface is shown with the white line. The light-gray line stands for the edge of the first Brillouin zone. The dark gray curves indicating surface waves are given for the eye. Different sorts of surface modes are designated with circles.
}
\end{figure*}

To characterize the dispersion of the surface waves we resorted to attenuated total internal reflection spectroscopy (see Fig.~\ref{fig:sample_setup}b). To excite the surface waves, one needs to provide excitation wavevectors residing under the light line of the dielectric environment of the metasurface (\textit{i.e.}, silicon oxide substrate and resist superlayer). For this purpose, we used a zinc selenide (ZnSe) hemicylindrical prism with the refractive index of around 2.48 in the NIR range\cite{Marple1964znse}. The sample was attached to the prism with a polymer screw  to minimize the weakly controllable air gap between the sample and ZnSe interfaces (see the inset in Fig.~\ref{fig:sample_setup}b). In this configuration (known as the Otto geometry), surface waves can be excited via evanescent coupling of light incident at the ZnSe-sample interface at angles greater than the critical angle, which is about 36$^\circ$ in the spectral range of interest. By measuring reflectance spectra at different angles of incidence $\theta$ (see Methods for the details of the experimental setup), we were able to reconstruct dispersion of surface waves excited at the metasurface.Numerical simulations mimicking the experiment were carried out using the frequency-domain solver of the COMSOL Multiphysics package (see Methods for details).

\section*{Discussion\label{sec:discussion}}

Figure~\ref{fig:exp_vs_num} shows the experimental and simulated reflectance maps plotted in ``wavelength - angle of incidence'' axes for both the TE- and TM-polarized excitations and three azimuthal angles $\varphi=0^\circ,45^\circ,90^\circ$ describing orientation of the plane of incidence with respect to the long axis of the nanodisks. The measured reflectance maps demonstrate full correspondence with the simulated ones for all considered cases. The insignificant discrepancy can be attributed to inhomogeneities of the sample and the weakly controllable air gap size in the experiment. 

The reflectance maps demonstrate a rich variety of near- and far-field features corresponding to the regions above and below the critical angle, respectively. The pronounced reflectance dip curves above the critical angle stand for the surface modes supported by the metasurface. In the NIR range ($\lambda > 750$~nm), two types of surface waves are excited: a short-wavelength TM-like plasmon and a long-wavelength TE-like plasmon \cite{sainidou2008plasmon,sun2014artificial}. Observation of these two types of plasmons agrees with the predictions of the local analytical model for resonant anisotropic metasurface~\cite{Yermakov2015}. The spectral position of the reflectance dips corresponding to these modes strongly depends on the orientation of the sample, clearly demonstrating the anisotropy of their dispersion. The TE-like mode has no frequency cut-off and can propagate at arbitrary low frequencies, where its dispersion curve asymptotically tends to the light line. This means that the mode becomes weakly confined and leaks into the ZnSe prism. It results in broadening of the resonance and decrease of its intensity.  The TM-like mode residing in the shorter wavelength region has a cutoff frequency that depends on the propagation direction. Importantly, due to controllable dispersion both TE- and TM-like plasmons can exhibit wavevectors and density of optical states larger than that of plasmons at gold-air interface. These properties are essential for high-resolution imaging and sensing.

The origin of the two-dimensional TE- and TM-like plasmons is the following. The TE-like mode is formed due to the coupling of electric dipoles induced in the plasmonic nanodisks in the direction perpendicular to the propagation direction of the wave. Existence of such mode is possible only for the negative polarizability of the plasmonic particles~\cite{Mikhailov2007}, i.e. at the frequencies lower than their plasmon resonance. The TM-like mode is formed due to the coupling of the dipoles oriented along the propagation direction. Existence of such mode is possible only for the positive polarizability of the plasmonic particles, i.e. at the frequencies higher than plasmon resonance. Due to the elliptic shape of the nanodisks and their interaction with neighbors, the degeneracy of the localized plasmon resonance is lifted. It results in the polarization hybridization of the 2D plasmons and pronounced anisotropy of their dispersion.

\begin{figure}[ht]
\includegraphics[width=1.0\linewidth]{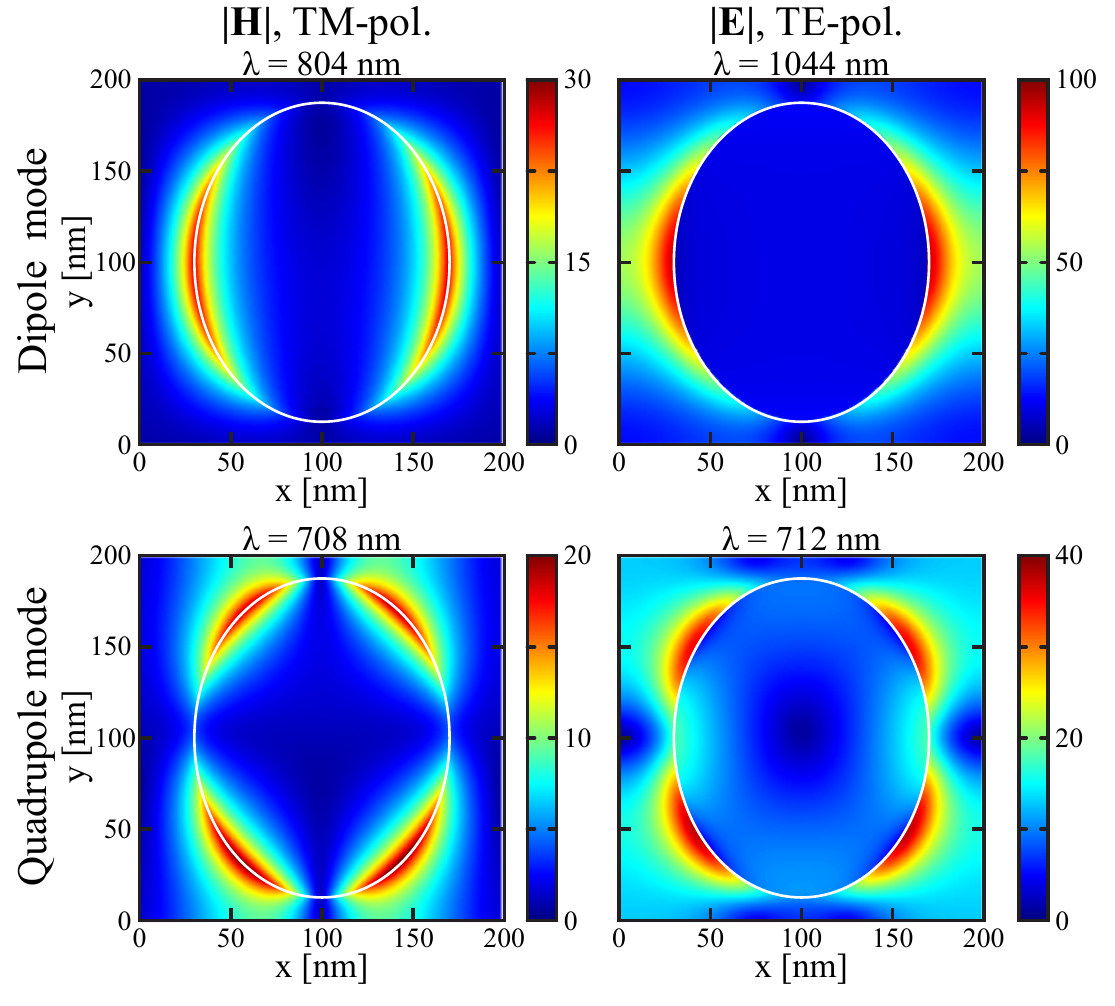}
\caption{
\label{fig:disksect}
Simulated magnetic and electric field distributions corresponding to the surface modes excited in the metasurface by TM- and TE-polarized waves. The plane of incidence is parallel to the long axis of elliptical particles. The incident  wavevector forms an angle of $\theta = 50$ degrees with surface normal. The respective excitation geometries are marked with circles in Fig.~\ref{fig:exp_vs_num}.
}
\end{figure}

In experiment, the dipole plasmon resonances are clearly manifested at small angles of incidence (below the critical angle) in both polarizations as the broad peaks. Strong dependence of their spectral position on orientation of the metasurface and polarization of the excitation wave confirms non-degeneracy of the dipole plasmon resonances of the nanodisks. The calculated field profiles for TE- and TM-like plasmons plotted in Fig.~\ref{fig:disksect} verify that the modes are formed due to coupling of the dipole plasmon resonances of the nanodisks.

The surface modes of plasmonic metasurfaces can be formed due to the coupling of higher multipole plasmon resonances, e.g., quadrupolar, octupolar etc~\cite{Pors2014}. So, the reflectance maps shown in Fig.~\ref{fig:exp_vs_num} contain additional branches near 700~nm, observed under both TM- and TE-polarized excitation. These modes are characterized by  barely visible dispersion (low group velocity) and a narrow spectral half-width both above and below the critical angle. The in-plane quadrupole modes in a thin nanodisk are dark modes for the normal excitation.
The quadropole nature of the modes becomes apparent from the field profiles (Fig.~\ref{fig:disksect}). Almost dispersionless behavior of the quadrupole modes is explained by stronger field localization and weaker interaction between neighboring particles in comparison with the dipole modes. 

\begin{figure*}[t!]
\includegraphics[width=1.0\linewidth]{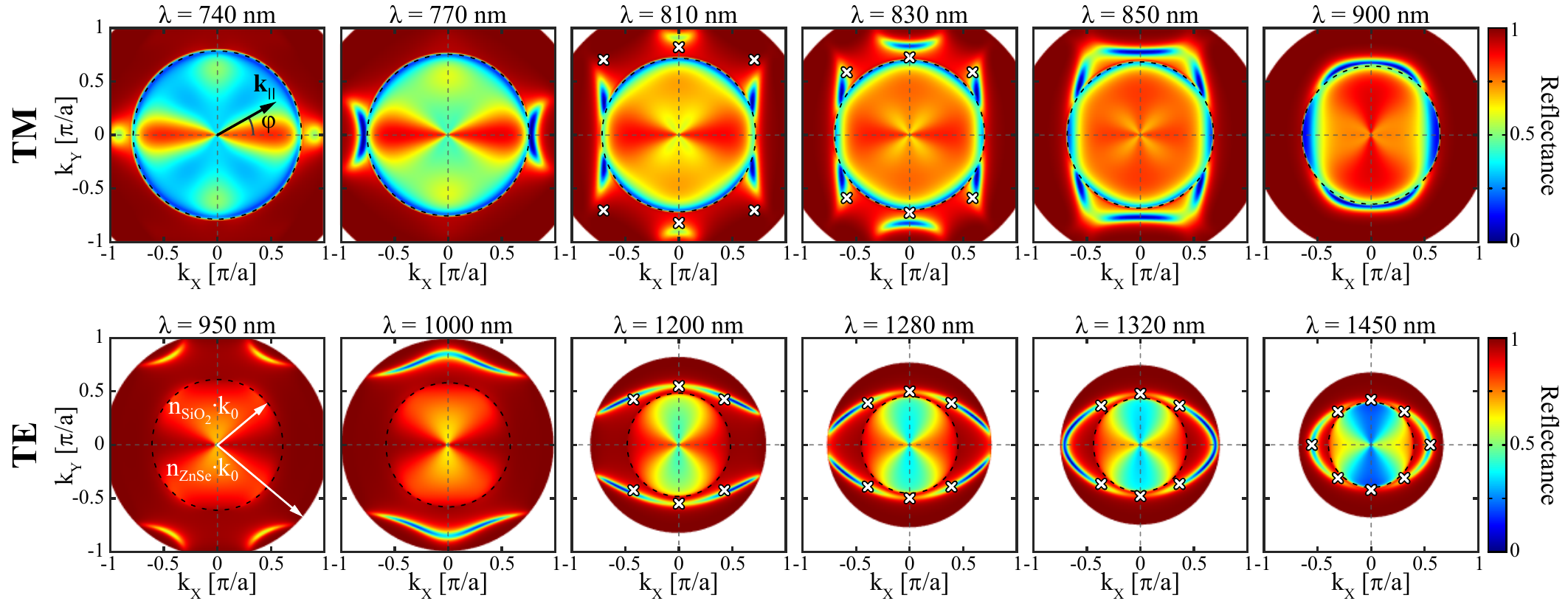}
\caption{
\label{fig:isofreq}
Reciprocal space reflectance maps for TM- (top row) and TE-polarized (bottom row) light demonstrating spectral evolution of isofrequency contours of surface waves. The maps are plotted within first Brillouin zone. The largest absolute value of available wavevectors (outer circular edge of the definition area) corresponds to the light circle in ZnSe: $k_{0}n_{ZnSe}$. The inner dashed black circle indicates light wavevector in fused silica substrate $k_{0}n_{SiO_2}$. The surface states reside between these circles. Crosses denote the experimental data.
}
\end{figure*}

Anisotropic properties of the 2D plasmons are manifested most clearly in isofrequency contours plotted within the first Brillouin zone. Figure~\ref{fig:isofreq}  demonstrates spectral evolution of the isofrequency contours calculated numerically for both TM- and TE-like surface waves. These contours are exhibited as the blue curves lying between the light circles of glass/resist and ZnSe. The respective experimental data are shown with crosses and demonstrate excellent agreement with numerical results. The minor discrepancies observed in the short-wavelength region are most likely to be concerned with the inaccuracy of the model of gold permittivity dispersion.

At low frequencies, far from the plasmon resonance, the anisotropy is vanishingly small, and an isofrequency contour of the TE-like plasmon is very close to a circle (not shown). With the increase of the frequency the circle clearly transforms into the ellipse ($\lambda = 1450$~nm). Then the contour ruptures giving rise to a forbidden range of propagation directions along the x-axis ($\lambda = 1280$~nm). At higher frequencies ($\lambda = 950$~nm) TE-like plasmon can propagate only in narrow angular bands in the vicinity of the diagonals of the Brillouin zone. Directivity of the TM-like plasmon (allowed angular band) demonstrates even more dramatic evolution with a wavelength. Near the frequency cutoff ($\lambda\approx900$~nm) the surface wave propagates nearly along the long axis of the nanodisks completely vanishing in the other directions. At higher frequencies ($\lambda\approx750$~nm) it propagates only along the short axis of the nanodisks. Such a tunability can be exploited for on-chip routing of optical signals.

The form of the isofrequency contours predicts the relationship between the group and phase velocities and defines the shape of the wavefront and character of propagation\cite{poddubny2013hyperbolic}. For example, a positive curvature corresponds to the divergent propagation, a negative curvature corresponds to the self-collimated propagation, a flat contour corresponds to the non-diffractive regime. So, the TM-like plasmon demonstrates self-focusing along the long axis of the nanodisks at $\lambda=830$~nm and along the short axis at $\lambda=770$~nm, when the hyperbolic regime takes place. Therefore, the considered metasurface support elliptic, hyperbolic, and more complex dispersion regimes, which could be fine-tailored at the fabrication stage by deliberate shaping of the particles.        

To conclude, we have studied dispersion and polarization properties of the surface states supported by a thin metasurface composed of resonant plasmonic nanoparticles. The particles are arranged in a dense square array with a subwavelength period. They have the shape of elliptic cylinders, and due to such shape the metasurface exhibits strong anisotropic properties. The metasurface has been characterized by means of total internal reflection spectroscopy in the broad range of wavelengths between 600 nm and 1600 nm under different angles of incidence and orientations of the structure. Existence of the resonant surface waves with anisotropic dispersion bands has been confirmed. The striking difference with the case of metal-dielectric interfaces
is that along with the expected TM-like plasmons the structure supports the TE-like surface states, both maintaining the highly directional propagation regime. Full-vectorial simulations consistently support our findings.

For both types of waves we have 
analyzed the spectral evolution of the isofrequency contours. Highly nontrivial and polarization-dependent transformation of the contours has been observed. Numerical analysis helped us to classify the principle bands as tightly bound electric dipole resonances. We have also shown that such metasurface supports quadrupole modes with extremely weak dispersion. Due to higher Q-factors of the quadrupole resonances the quadrupole surface waves are characterized by weaker interparticle coupling and thus very flat dispersion. Our findings on a resonant anisotropic metasurface supporting directional and polarization-dependent surface waves provide a flexible platform for on-chip surface photonics for various applications, such as processing and routing of optical signals in quantum communication systems, high-resolution sensing and enhancement of non-linear processes.

\section*{Methods.\label{sec:methods}}
\subsection*{Experimental investigations.\label{subsec:methods_exp}}
The anisotropic metasurface was fabricated on a glass substrate using electron beam lithography followed by thermal evaporation of a 20~nm thick layer of gold and a lift-off process. 
The fabricated structure is a 200$\times$200~$\mu$m$^2$ array of cylindrical gold nanodisks with the elliptical base. The period of the array is 200~nm, while the long and short axes of the nanodisks are 175 and 140~nm, respectively (see Fig.~\ref{fig:sample_setup}a).

In the experiment, the sample was illuminated by a supercontinuum laser source (NKT Photonics SuperK Extreme) polarized with a Glan-Taylor prism and focused by a series of parabolic mirrors on the sample surface through a ZnSe prism to a spot of approximately 150~$\mu$m in size. The reflected light was collected with another parabolic mirror and then sent to a spectrometer (Ando AQ-6315E) through an optical fiber. The sample and the collection optics were mounted on separate rotation stages, which allowed for reflectance spectra measurements in a broad angular range (from 10$^\circ$ up to 60$^\circ$).

\subsection*{Numerical modelling.\label{subsec:methods_sim}}
Numerical simulations were carried out using the frequency-domain solver of the COMSOL Multiphysics package. The simulation cell with periodic boundary conditions in both directions is shown schematically in Fig.~\ref{fig:sample_setup}b. The exact dimensions of the structure were verified by means of scanning electron microscopy (Fig.~\ref{fig:sample_setup}a). The refractive indices of the materials were obtained from literature (ZEP 520A\cite{Zep520}, zinc selenide\cite{Marple1964znse}, gold\cite{McPeak2015gold} and fused silica\cite{Malitson1965sio2}). The size of the air gap in the simulations was chosen to be 25~nm accordingly to the best matching of the simulated spectra with the experimental data.

\section*{Acknowledgements}
O.T., R.M. and A.V.L. acknowledge partial support from the Villum Fonden through "DarkSILD" project No. 11116. 
The reflectivity measurements and the theoretical part of this work were financially supported by Russian Science Foundation (No. 15-12-20028).

\section*{Author contributions}

\subsection*{Contributions} 
I.V.I. and A.A.B. conceived the idea.
A.S., I.M. and O.T. conducted the experiments.
D.B. and O.Y. performed the sample design optimization. 
R.M. fabricated the samples. 
D.V.P. performed the numerical simulations. 
A.S., I.S.S. and A.A.B. wrote the manuscript.
A.V.L. supervised the project.
All authors contributed to discussions and proofreading of the manuscript.

\subsection*{Corresponding author}
Correspondence and requests for materials should be addressed to Anton Samusev 
(email: \url{a.samusev@metalab.ifmo.ru}).



\section*{Competing financial interests}
Competing financial interests: The authors declare no competing financial interests.





\end{document}